\documentclass[aps, pre,amsmath,amssymb,twocolumn,floatfix]{revtex4}
\usepackage{graphicx}
\input epsf

\begin{document}

 \title{\bf Metastable liquid-liquid coexistence and density 
anomalies in a core-softened fluid}

\author{H.M. Gibson}
\author{N.B. Wilding}
\address{Department of Physics, University of Bath, Bath BA2 7AY, United Kingdom}

\tighten

\begin{abstract} 

Linearly-sloped or `ramp' potentials belong to a class of core-softened
models which possess a liquid-liquid critical point (LLCP) in addition
to the usual liquid-gas critical point. Furthermore they exhibit
thermodynamic anomalies in the density and compressibility, the nature
of which may be akin to those occurring in water. Previous simulation
studies of ramp potentials have focused on just one functional form, for
which the LLCP is thermodynamically stable. In this work we construct a
series of ramp potentials, which interpolate between this previously
studied form and a ramp-based approximation to the Lennard-Jones (LJ) 
potential. By means of Monte Carlo simulation, we locate the LLCP, the
first order high density liquid (HDL)-low density liquid (LDL)
coexistence line, and the line of density maxima for a selection of
potentials in the series. We observe that as the LJ limit is approached,
the LLCP becomes metastable with respect to freezing into a hexagonal
close packed crystalline solid. The qualitative nature of the phase
behaviour in this regime shows a remarkable resemblance to that seen in
simulation studies of accurate water models. Specifically, the density
of the liquid phase exceeds that of the solid; the gradient of the
metastable LDL-HDL line is negative in the pressure ($p$)-temperature ($T$)
plane; while the line of density maxima in the $p$-$T$ plane has a shape
similar to that seen in water and extends well into the {\em stable}
liquid region of the phase diagram. As such, our results lend weight to
the `second critical point' hypothesis as an explanation for the
anomalous behaviour of water.

\end{abstract} 
\maketitle 
\epsfclipon  


\section{Introduction}

Multiple liquid phases are a common feature of the phase diagrams of
multi-component mixtures \cite{ROWLINSON}. However, there is a growing
body of experimental and computational evidence to indicate that they
can also occur in {\em single component} systems. Examples of such have,
to date, been found in a number of elemental systems including ({\em
inter alia}) Sulphur \cite{ANGILELLA03,BRAZHKIN92,TSE99}, Phosporous
\cite{KATAYAMA02A,KATAYAMA02B,HOHL92}, Hydrogen
\cite{SCANDOLO02,BONEV04,PFAFFENZELLER97} and Selenium
\cite{KATAYAMA01}. Additionally, tentative evidence has recently emerged
for the existence of liquid-liquid transitions in molecular liquids such
as n-butanol and triphenyl phosphite \cite{KURITA05}.

Arguably, however, the most intriguing example of a molecular system
exhibiting signs of a liquid-liquid (LL) transition, is water. Here the
``second critical point'' hypothesis \cite{POOLE} proposes that the LL
transition is wholly metastable with respect to freezing and that the
associated liquid-liquid critical point (LLCP) is responsible for the
celebrated thermodynamic anomalies in the density and compressibility in
the stable and metastable liquid region near the freezing boundary.
Support for this proposal comes from molecular dynamics simulations of
the (generally successful) TIP5P interaction potential \cite{YAMADA02}.
These find a metastable LL transition and associated critical point,
with a line of density maxima which seems to emanate from near the LLCP.
Moreover, it has been suggested that at very low temperature the LL
boundary evolves into a transition between low density and high density
glassy phases. Whilst transitions between amorphous phases of different
densities have been observed experimentally, their relationship to the
liquid phases is still a matter of some debate (see eg.
\cite{MISHIMA02,GIOVAMBATTISTA}). To date, however, the LL transition
has not been observed in real water, possibly because the metastable
lifetime of these phases is too small to be resolvable experimentally.

Notwithstanding the progress in identifying and characterizing LL phase
transitions and thermodynamic anomalies across a variety of disparate
systems, it remains unclear as to whether these phenomena are
pluralistic in physical origin, or can instead be traced to a single
common underlying mechanism. Furthermore, the connection between LL
transitions and thermodynamic anomalies seems at present rather tenuous:
in some systems density maxima have been reported without (as yet)
evidence of a LL transition, as is the case in SiO$_2$ \cite{ANGEL76};
while in other (indeed in {\em most}) systems for which LL transitions
have been reported, they appear unaccompanied by anomalies (see eg.
\cite{ANGILELLA03,MCMILLAN04} for reviews). Only in water do the two
phenomena seem fairly firmly linked. It is therefore of interest to
enquire as to whether there exist simple model potentials that captures
the general qualitative behaviour of a LL transition and thermodynamic
anomalies, and to elucidate their properties in detail.

Work in this direction has concentrated on the so-called core-softened
potentials, originally introduced by Hemmer and Stell \cite{STELL}. The
functional form of these potentials is engineered to favor two distinct
interparticle separations --thereby providing impetus for a transition
between two liquids of differing densities.  Core-softened potentials
can usefully be subdivided into two classes: ``shoulder'' potentials in
which the hard core exhibits a region of negative curvature, and
``ramp'' potentials in which the hard core is softened via a linear
slope. To date, the majority of work on core softened potential has
concentrated on the shoulder form.  Simulation and theory
\cite{PELLICANE,FRANZESE01,QUIGLEY,MAUSBACH,MALESCIO05} show that these
do indeed (given a favorable choices of potential parameters) possess a
metastable LLCP. However, to date, no firm evidence of thermodynamic
anomalies has been reported. (Initial indications of anomalies in 2D
shoulder models \cite{SL98} were subsequently shown \cite{WILDING02} to
be an artifact of the quasi-continuous nature of the 2d freezing
transition in the case when the solid has a density less than the
liquid.)

In contrast to their shouldered counterparts, ramp potentials are known
to exhibit both a LLCP and thermodynamic anomalies. They therefore
appear a potentially more fruitful route to determining whether the
qualitative features of the LL transition in systems such as water can
be described by a very simple model, as well as for investigating the
more general aspects of the relationship between the LL phase behaviour
and thermodynamic anomalies. As we shall show in the present paper, a
ramp model can indeed capture (to a remarkable degree) the qualitative
features of the metastable LL transition and density maxima seen in
simulations of accurate water models.

\section{Ramp potentials}

The phase behaviour of particles interacting via an isotropic pair
potentials in which the steep repulsive core is softened by a linear
ramp, was first considered 35 years ago by Hemmer and Stell
\cite{STELL}. Their calculations for a one-dimensional system revealed a
range of parameters for which two phase transitions occurred, and they
speculated that the same might be true in high dimensions. More
recently, interest in such potentials has been rekindled following
computer simulation and mean field studies of ramp potentials in two and
three dimension by Jagla \cite{JAGLA01}. These revealed evidence of HDL
and LDL phases in addition to the expected liquid and gas phases, and
the presence of density and compressibility anomalies. A subsequent
detailed MC simulation study of the same system by one of us
\cite{WILDING02}, accurately mapped a portion of the HDL-LDL phase
boundary, the liquid-gas boundary and the locus of the lines of density
and compressibility maxima.

The form of ramp potentials we consider in the present work is given by:

\begin{eqnarray}
\label{equation}
U(r) &=& \infty \:; ~~~~ r < r_0 \\\nonumber
U(r) &=& \frac{(r_0 - r)(D + 0.69)}{(r_1 - r_0)} + 0.69\:;~~~~ r_0 \le r < r_1 \\\nonumber
U(r) &=& D (r_2 - r)/(r_1 - r_2)\:;~~~ r_1 \le r < r_2 \\\nonumber
U(r) &=& 0\:; ~~~~ r > r_2\:,
\end{eqnarray}
where $U(r)$ is measured in units of $k_BT$. Given a constant hard-core
radius $r_0$ and contact value $U(r\to r_0^+)$, the form of the potential is determined by the position of
the minimum $r_1$, the maximum range of the potential $r_2$, and the
magnitude of the well depth $D=-U(r_1)$. In the original work of
refs.~\cite{JAGLA01,WILDING02}, this potential was studied for the
parameters values $r_1=1.72\:r_0,\;r_2=3.0\:r_0,\; D=0.1984$. In the present
work we study the properties of a {\em family} of such potentials, the
members of which are chosen such as interpolate between the original form and a ramp potential approximation
to the LJ potential. The interpolation simultaneously reduces the radius
of the potential minimum and the maximum range, whilst increasing the
potential depth. This is done in such a way as to maintain approximate
constancy of the second virial coefficient, thereby ensuring that the
potentials are comparable in a corresponding states sense
\cite{VLIEGENTHART00,NORO00}.

\begin{figure}[h]
\includegraphics[width=8.5cm,clip=true]{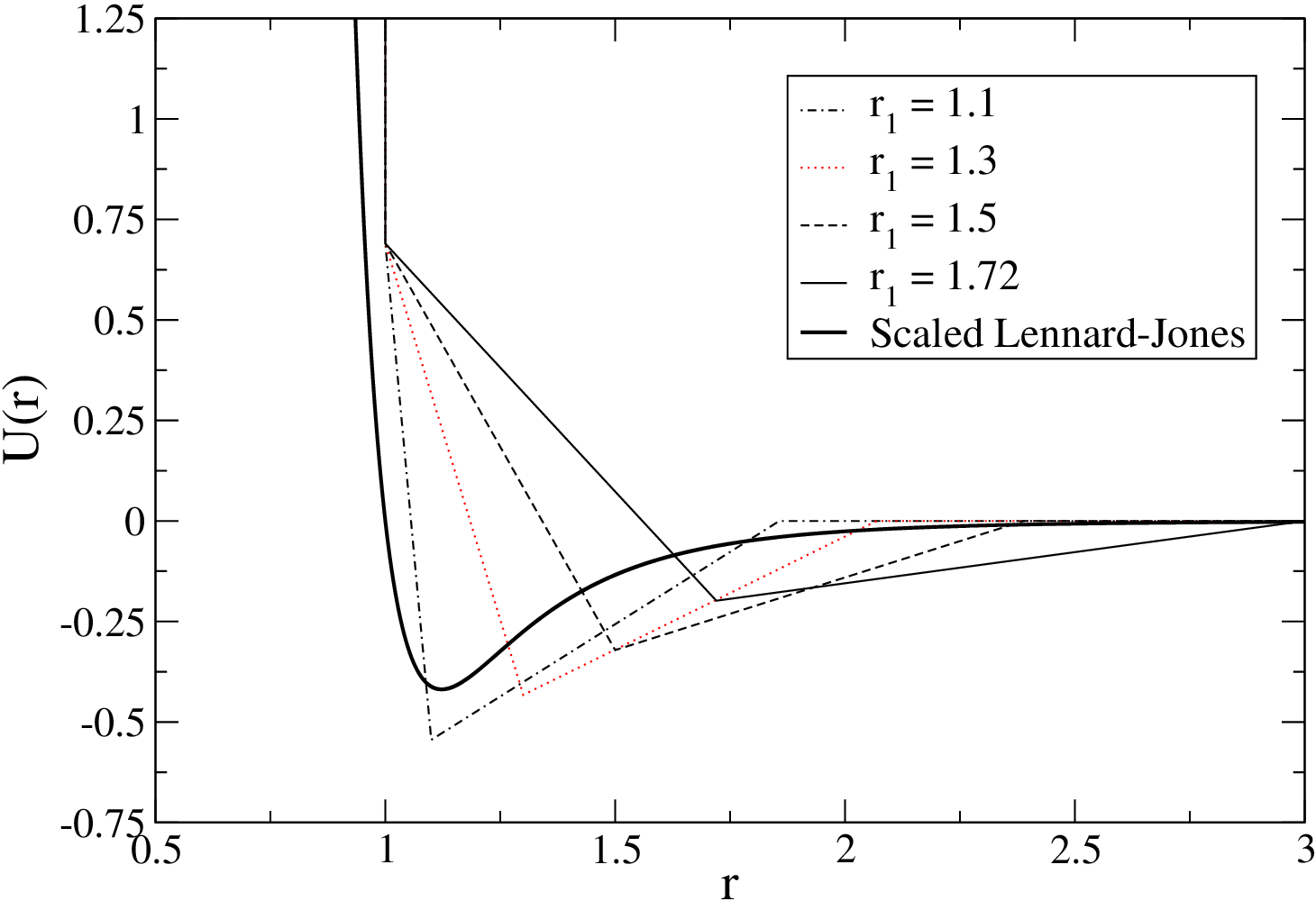}
\caption{A selection from the family of ramp potentials studied in this
work and listed in tab.\ref{tab:pots}. Also shown for comparison is the
LJ potential whose well depth has been scaled such that the second
virial coefficient takes the value $B_2=1.52$.}
\label{fig:family}
\end{figure}

We define our family of ramp potentials as follows. The limiting value
of the potential at the hard core contact is held constant at $U(r\to
r_0^+)=0.69$. Choosing to label each member of the family
of potentials by the radius of the minimum $r_1$, the associated well
depth $D(r_1)$ is given by $D(r_1)=1.1578-0.5578r_1$, while the value
of $r_2$ is tuned such to maintain the second virial coefficient at the
value $B_2=1.52$. The resulting values of $D(r_1)$ and $r_2$ are listed
in tab.~\ref{tab:pots}, and a selection of potentials is shown in
fig.~\ref{fig:family}.

\begin{table}
\begin{tabular}{|l@{\hspace*{5mm}}l@{\hspace*{5mm}}l@{\hspace*{5mm}}|}
\hline
$r_1$ & $D(r_1)$ & $r_2$ \\ \hline
1.72   &     0.1984   &      3.0 \\
1.7    &    0.209556  &      2.92483 \\
1.68   &     0.220711 &      2.8549 \\
1.66   &     0.231867 &      2.78963 \\
1.65   &     0.237444 &      2.75859 \\
1.64   &     0.243022 &      2.72853 \\
1.63   &     0.2486   &      2.69941 \\
1.62   &     0.254178 &      2.67118 \\
1.61   &     0.259756 &      2.6438 \\
1.60   &    0.265333  &      2.61721 \\
1.58   &    0.276489  &      2.56631 \\
1.5    &    0.321111  &      2.38847 \\
1.4    &    0.376889  &      2.21068 \\
1.3    &    0.432667  &      2.06848 \\ \hline
\end{tabular}
\caption{Forms of the ramp potentials studied in this work (cf. fig.~\ref{fig:family}). The well
depth is given by $D(r_1)=-U(r_1)=1.11578-0.568$, while the maximum
range $r_2$ is tuned such to maintain the second virial coefficient at
the value $B_2\approx 1.52$.}
\label{tab:pots}
\end{table}

\section{Computational Methods}

Monte Carlo simulation studies of the phase behaviour of a selection of
potentials in the family have been performed within the constant-$NpT$
ensemble. In the results reported below, temperature ($T$) is measured
in units of the well depth, particle number density ($\rho$) is measured
in units of $r_0^3$ and the pressure ($p$) in units of $k_BT/r_0^3$. All
simulations were performed for systems of $N=300$ particles. 

The principal aim was to locate the liquid-liquid (LL) coexistence line
and critical point for each potential and to map the line of density
maxima.  The coexistence boundaries were obtained using multicanonical
Monte Carlo techniques and histogram extrapolation in the well
established manner \cite{BERG92,FERRENBERG,WILDING01,BRUCE03}. To
estimate critical parameters, we have employed a crude version of the
finite-size scaling (FSS) analysis described in ref.~\cite{WILDING95}.
The analysis involves scanning the range of pressure $p$ and temperature
$T$ until the observed probability distribution of the fluctuating
instantaneous particles density  matches the independently known
universal fixed point form appropriate to the Ising universality class
in the FSS limit. Owing to the relatively small temperature at which the
liquid-liquid critical point generally occurs, the acceptance rate for volume
updates in the constant-$NpT$ ensemble are very low, resulting in
extended correlation time for the density fluctuations. Consequently we
were able neither to study a wide range of system sizes nor obtain data
of sufficient statistical quality to permit a more sophisticated FSS
analysis. Nevertheless it transpires that our estimated uncertainties
for the critical point parameters are sufficient to resolve clearly the
trends that occur as the form of the potential is altered. 

Lines of density maxima were mapped by measuring the density as a
function of temperature along isobars of the phase diagram. The task of
tracking the line of maxima was again aided by histogram extrapolation
techniques. 

\section{Results}

\subsection{The LDL-HDL transition and the density anomaly}

\begin{figure}[h]
\includegraphics[width=8.0cm,clip=true]{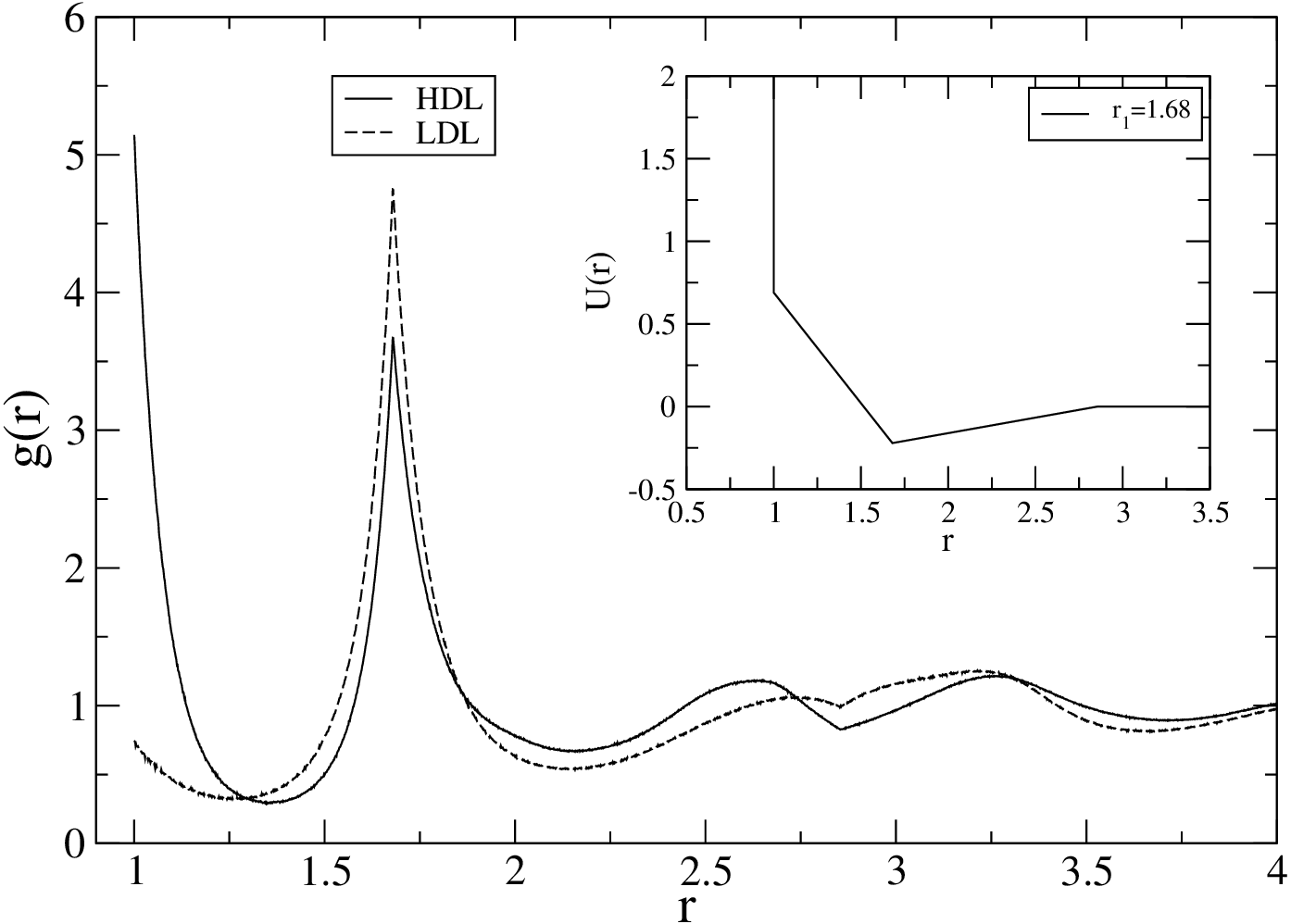}
\caption{Comparison of the forms of $g(r)$ for coexisting points on the
LL phase boundary. Parameters are $r_1=1.68$, $T=0.0644=0.944T_c$,
$p=0.05021$. The density of the HDL phase is $\rho=0.484(1)$, while that of
the LDL phase is $\rho=0.313(1)$. }
\label{fig:g_r}
\end{figure}

It is natural to enquire, first of all, as to the structural differences
between the LDL and HDL phases. In fig.~\ref{fig:g_r}, we show the form
of the radial distribution function $g(r)$ at a LL coexistence point for
the $r_1=1.68$ potential. The temperature is some $5\%$ below that of
the LLCP. From the figure, it is evident that in the LDL phase, the  majority of
first neighbors are located at the potential minimum, whereas in
the HDL phase, there is a much larger number of neighbors at the hard core
diameter and fewer at the potential minimum.

\begin{figure}[h]
\includegraphics[width=8.5cm,clip=true]{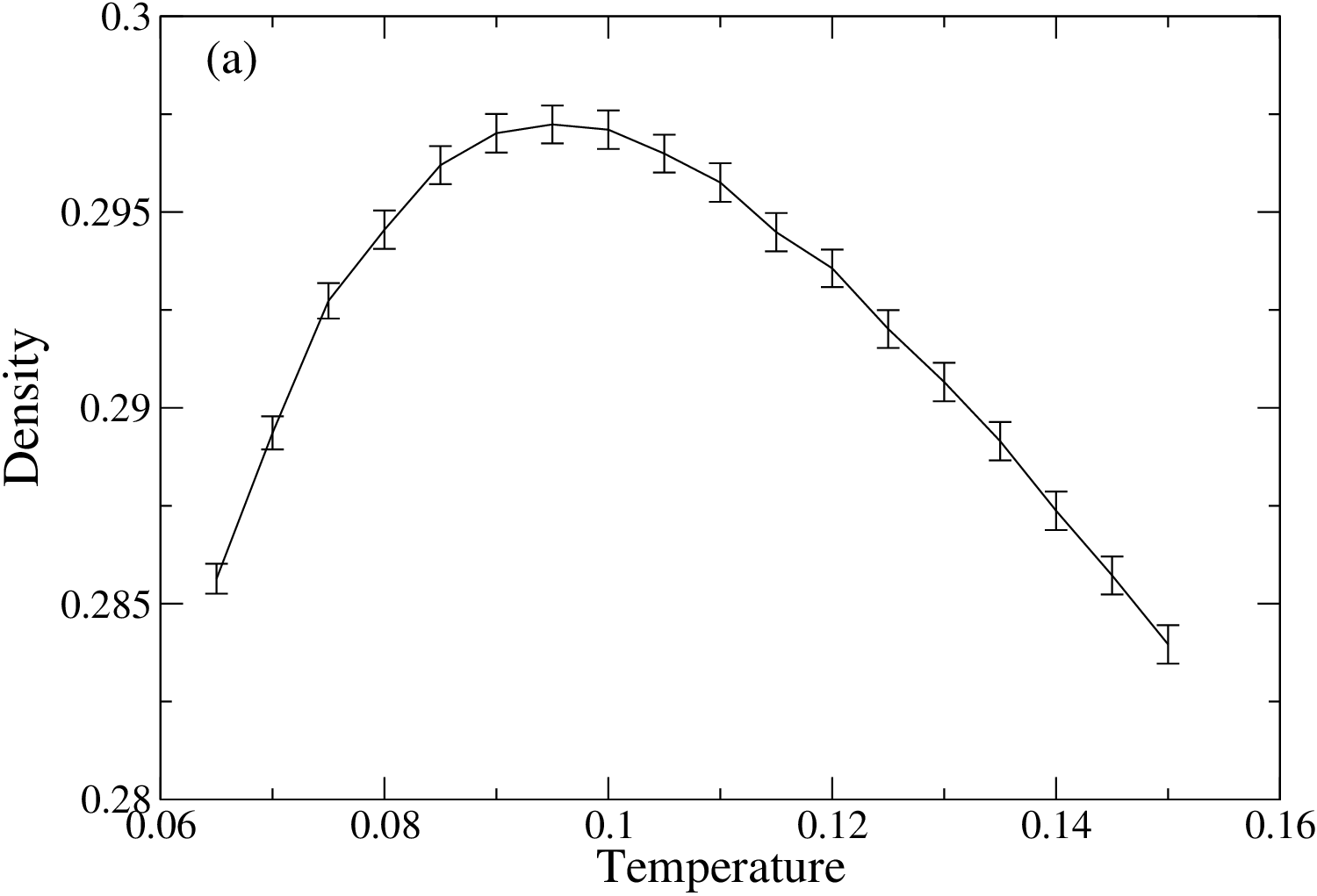}
\includegraphics[width=8.0cm,clip=true]{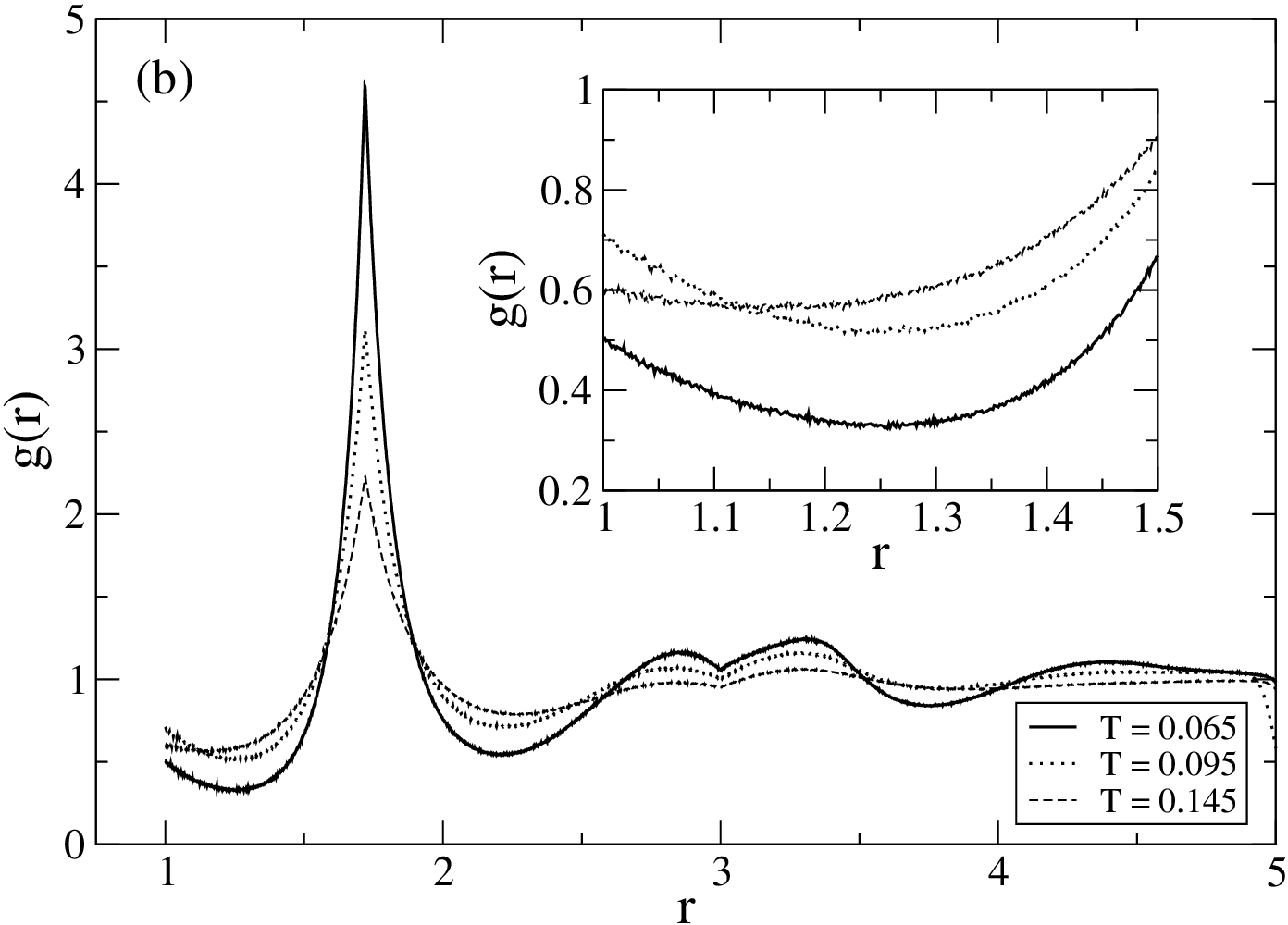}
\caption{{\bf (a)} The measured number density as a function of
temperature for $p=0.0247$, for the potential having $r_1=1.72$.  {\bf
(b)} The measured form of the radial distribution function $g(r)$ for
the same potential at three temperatures on the $p=0.0247$ isobar spanning 
$T_{MD}=0.095(5)$.}
\label{fig:den_an}
\end{figure}

In refs.~\cite{JAGLA01,WILDING02} it was shown that the ramp potential
exhibits a maximum in the density as the temperature is lowered
isobarically though the LDL phase. An example is shown in
fig.~\ref{fig:den_an}(a) for the case $r_1=1.72$ at $p=0.0247$. For
these parameters the density maximum occurs at $T_{\rm MD}=0.095(5)$.
Also included in this figure is the form of the radial distribution
function $g(r)$ for three temperatures on this isobar: one below, one
above, and one at $T_{\rm MD}$. At the hard core contact value, one sees
that $g(r_0)$ is greatest for $T=T_{\rm MD}$. Thus the anomalous density
increase with increasing $T$ is apparently due, in part at least, to an
increase in the number of particles choosing the shorter of the two separation
distances and settling at the hard core, despite the higher energy cost.

We have traced the locus of the density maxima in the $p$-$T$ plane for
several of the potentials studied. These are discussed in connection with
the phase diagram in the following subsection. 

\subsection{Phase behaviour}

\begin{figure}[h]
\includegraphics[width=8.5cm,clip=true]{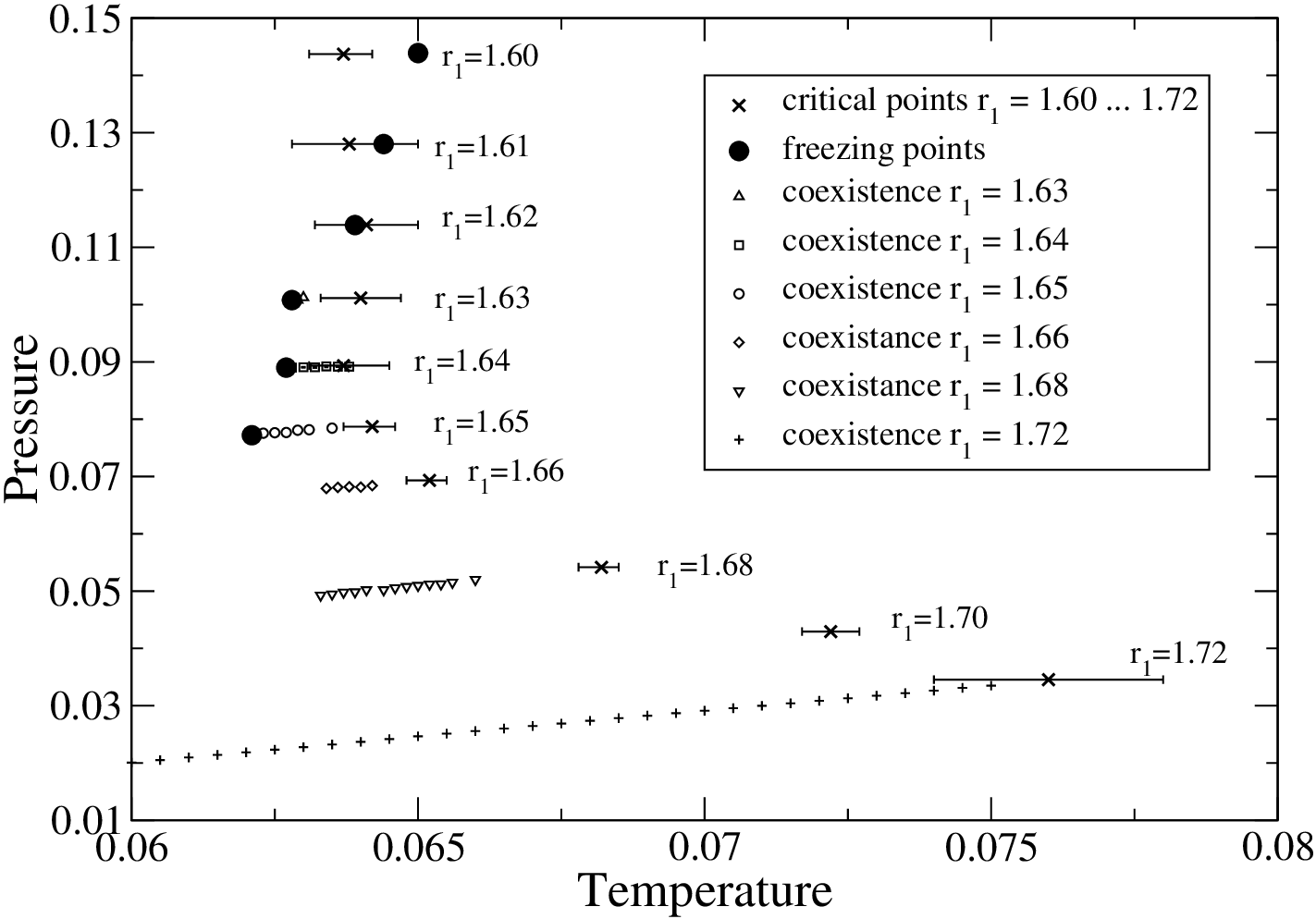}
\caption{The near-critical region of the phase diagram for each of the
ramp potentials studied. Shown in each instance is the estimated
location of the LLCP. For certain larger values of $r_1$ in the range
studied, a segment of the LDL-HDL phase boundary has also been
estimated. The point of intersection of the LL phase boundary with the
freezing line is shown for potentials in which the LLCP is either
metastable or only moderately stable with respect to freezing. Error
bars  represent the uncertainties in the critical temperature.
Uncertainties in the critical pressure, as well as in the location of
the LL line and the freezing points are comparable with the symbol
sizes.}
\label{fig:pd}
\end{figure}

Fig.~\ref{fig:pd} shows the location of the LLCP  for a selection of
values of $r_1$, together with (in some instances) a portion of the
associated LL coexistence line. One sees that as $r_1$ is decreased, the
LLCP shifts to lower temperatures and higher pressures. On tracking the
LL boundary down in temperature from the critical point, we observed
spontaneous freezing to a hexagonal close packed (hcp) structure. This
solid has a density lower than that of either liquid phase \cite{NOTE0}

The freezing point on the LL boundary shifts to higher temperatures as
$r_1$ is decreased. This, coupled with the concomitant decrease in the
critical temperature, rapidly narrows the temperature range over which
the LL transition is stable as $r_1$ is reduced. For $r_1\lesssim 1.62$,
the critical point became metastable with respect to the stable hcp
solid phase. For $r_1=1.61$ and $r_1=1.60$, simulations initiated in the
liquid phase was able to sample the near critical point fluctuations for
a limited time, before eventually spontaneous crystallization occurred.
Freezing occurred very rapidly in the critical region for $r_1<1.60$,
preventing accurate estimates of critical point parameters or indeed the
value of $r_1$  at which the critical point becomes completely unstable
rather than simply metastable.

\begin{figure}[h]
\includegraphics[width=8.0cm,clip=true]{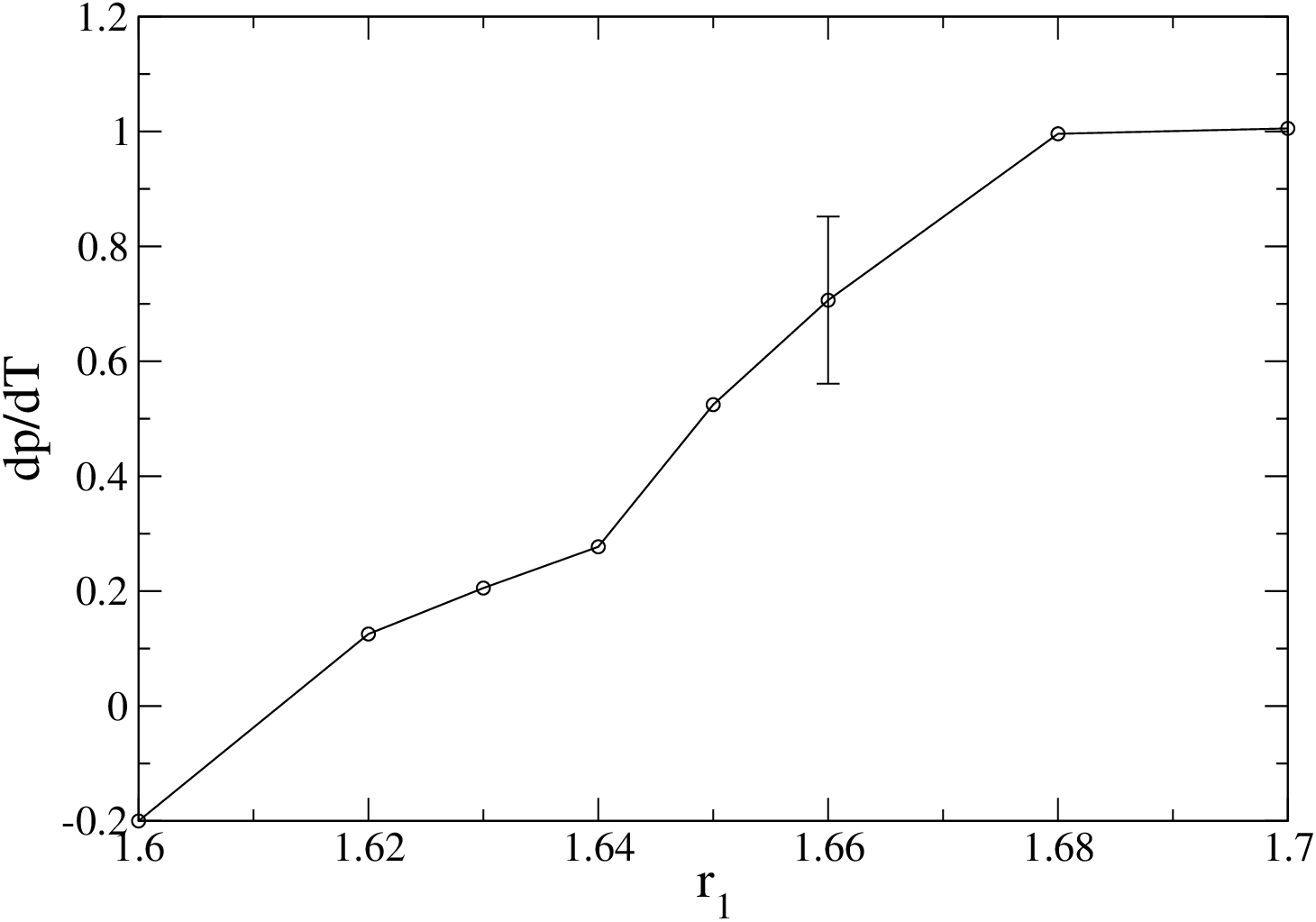}
\caption{Estimates of the near-critical gradient of the LDL-HDL
coexistence boundary in the $p-T$ plane, for the family of potentials
shown in \protect\ref{fig:family}. A representative error bar is shown.} 
\label{fig:gradient} 
\end{figure}

It is interesting to note that as $r_1$ is decreased, the initially positive
gradient of the LDL-HDL line in the $p-T$ plane reduces in magnitude and
changes sign close to the point at which the LDL-HDL critical point
becomes metastable. This trend is quantified in fig.~\ref{fig:gradient}.
Thus the gradient of the {\em metastable} LDL-HDL line is negative, as
has also been reported to be the case for water \cite{SCALA00}. We find
that the change in the sign of the gradient occurs because the enthalpy
difference between the two phases changes sign, rather than the density
difference.
 
\begin{figure}[h]
\includegraphics[width=8.5cm,clip=true]{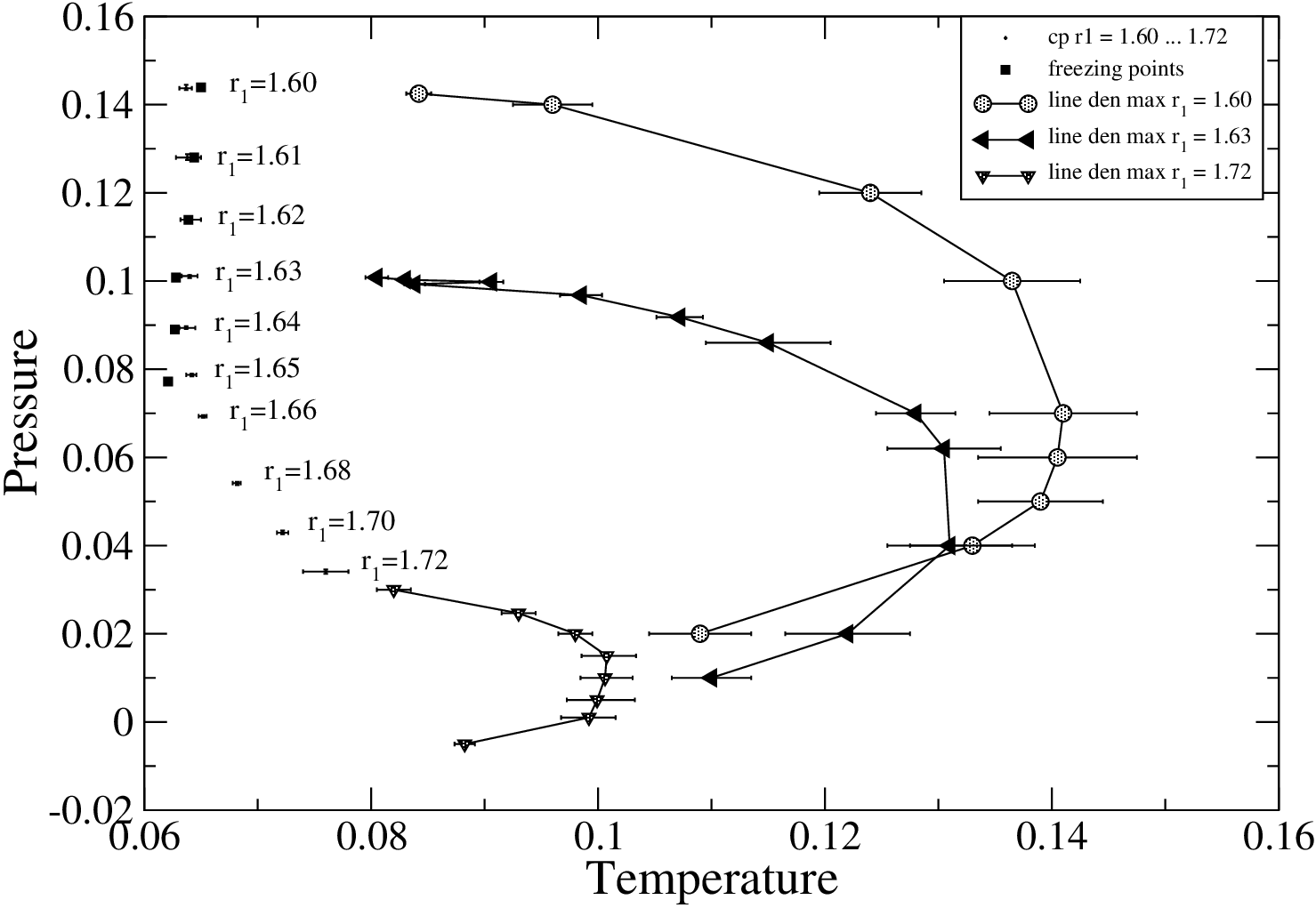}
\caption{Lines of density maxima for a selection of the potentials studied,
superimposed upon the phase diagrams of fig.~\ref{fig:pd}.} 
\label{fig:anomalies} 
\end{figure}

Fig.~\ref{fig:anomalies} superimposes the lines of density maxima on the
phase diagrams of several of the ramp potentials studied. The gradient
of these lines changes from negative to positive values in the $p$-$T$
plane as pressure is reduced. We observe a strong increase in the temperature
of the turning point as $r_1$ decreases. It is noteworthy that the
shape of the line of density maxima is similar to that found in MD
simulations of TIP5P water \cite{YAMADA02}. Furthermore, for the case
$r_1=1.60$, for which the LDL-HDL critical point is significantly buried
within the stable solid region, the density anomaly is nevertheless
observable over a wide range of the stable liquid phase. 

\begin{figure}[h]
\includegraphics[width=8.0cm,clip=true]{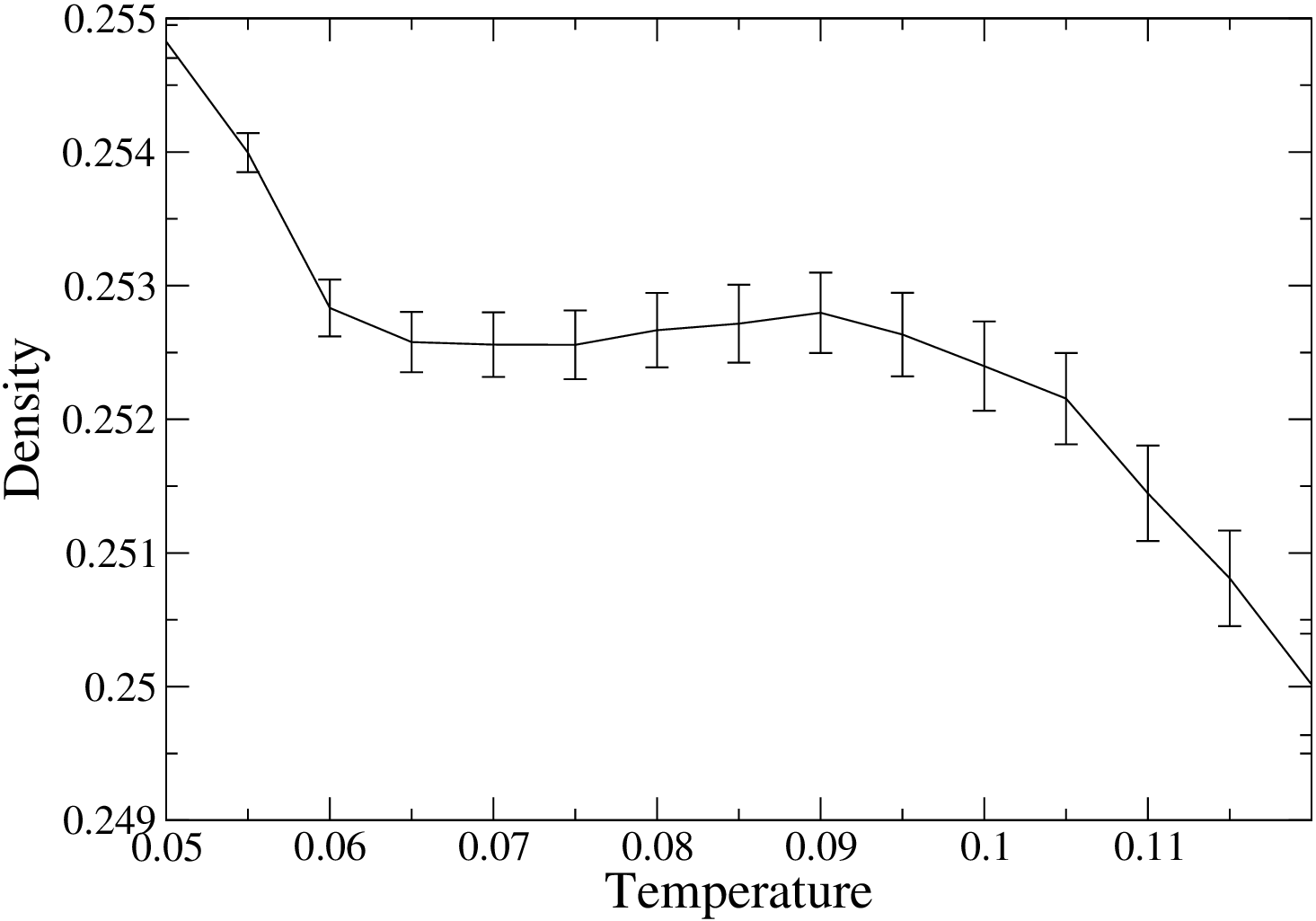}
\caption{The measured number density as a function of
temperature at $p=-0.005$, for the potential having $r_1=1.72$. The
system freezes as $T=0.055$.} 
\label{fig:den_min} 
\end{figure}

At high pressure, the lines of anomalies for the various potentials
becomes rather flat and appear to approach the respective LLCP. At low
pressure we find that the line is truncated by freezing to a face
centered cubic (fcc) solid structure at slightly negative pressures. For
most of the potentials studied, the anomalous decrease in density
continues right up to the stable solid region as $T$ is lowered isobarically; there is 
no subsequent density minimum, i.e. a return to ``normal'' behaviour, as
has been recently reported in simulations of the ST2 model of water
\cite{POOLE05}. An exception is the the case of $r_1=1.72$ at small negative
pressure. Here the density maximum flattens and a small
minimum appears, followed by a clear increase in the
density as freezing is approached, as is shown in fig.~\ref{fig:den_min}.

\begin{figure}[h]
\includegraphics[width=8.0cm,clip=true]{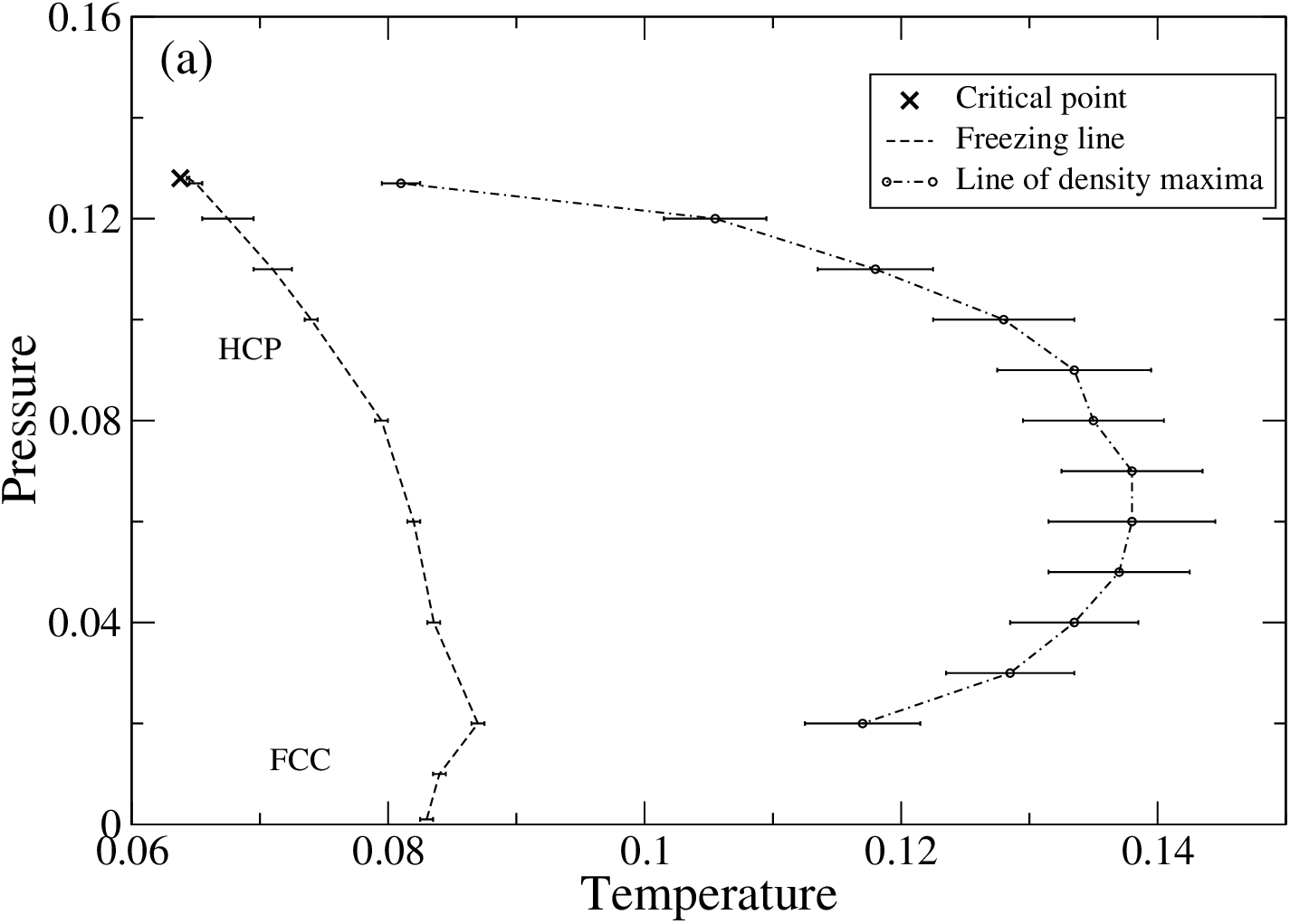}
\includegraphics[width=8.5cm,clip=true]{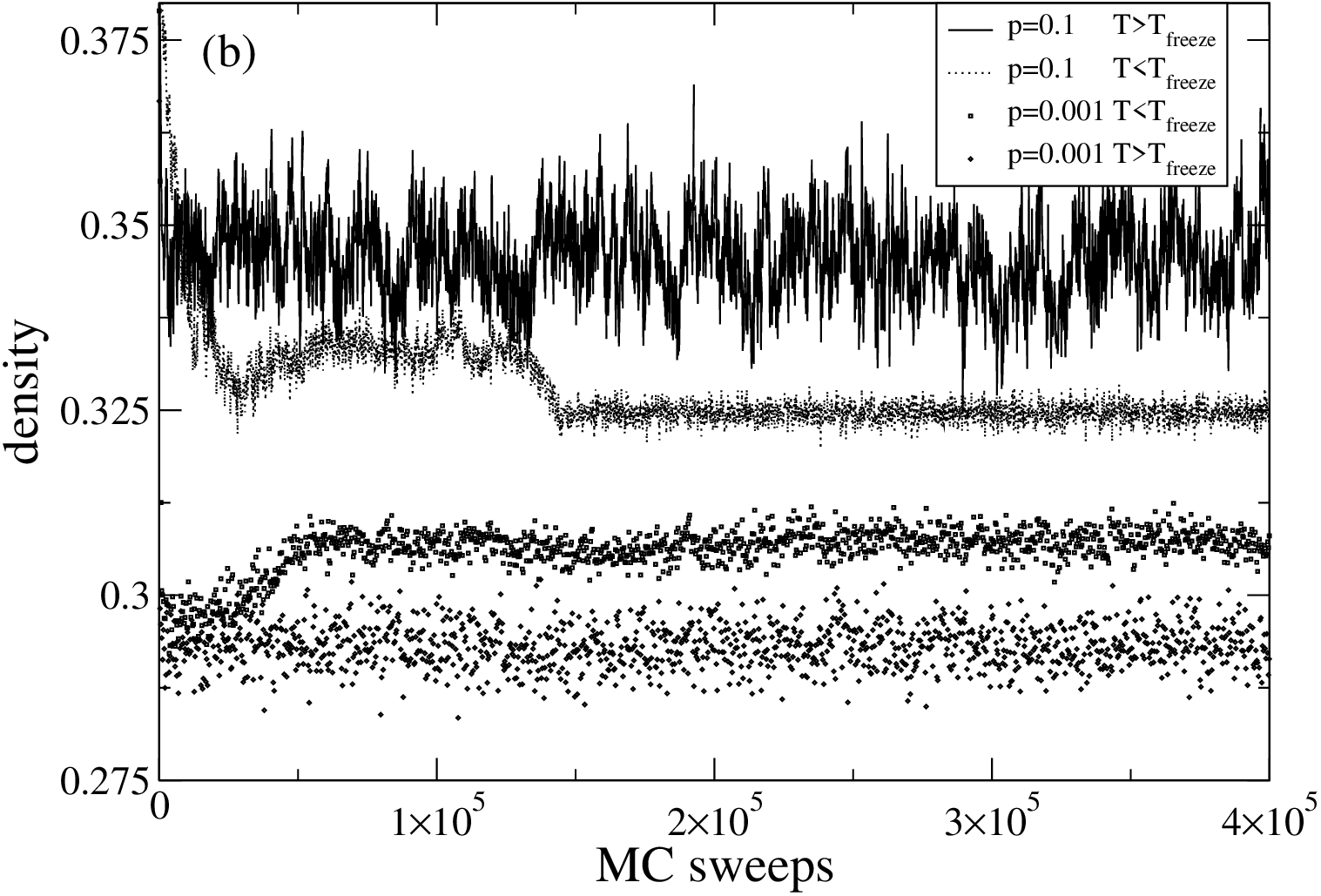}
\caption{{\bf(a)} The liquid-solid coexistence boundary and line of
anomalies in the $p$-$T$ plane for $r_1=1.61$. {\bf(b)} The time
evolution of the system density close to the freezing transition, as
described in the text. The figure shows the freezing to a solid of
higher (lower) density for low (high) pressures respectively. Time
is measured in units of Monte Carlo sweeps.} 
\label{fig:freezeline} 
\end{figure}

We have estimated the locus of the liquid-solid coexistence boundary in
the $p$-$T$ plane for the potential having $r_1=1.61$, for which
the LLCP is metastable; the results are presented in
fig.~\ref{fig:freezeline}(a).  Fig.~\ref{fig:freezeline}(b) shows for
both a high and low pressure, the time evolution of the simulation
density starting from an initial liquid-like configuration for two
temperatures either side of the freezing point. In the case of the
higher pressure ($p=0.1$), the system freezes to a hcp solid of lower
density, while for lower pressure ($p=0.001$), the solid is fcc in
structure having a density greater than that of the liquid. One thus
expects that the gradient of the freezing boundary in the $p-T$ plane is
negative at high pressure and positive at low pressure. This is indeed
confirmed by fig.~\ref{fig:freezeline}(a): within the rather limited
accuracy of our measurements, the gradient of the freezing boundary
appears to change sign at $p\simeq 0.02$, suggesting that this marks a
triple point between hcp, fcc and liquid phases. We have not attempted
to map the hcp-fcc coexistence line within the solid region, although on
cooling the fcc structure, we find it transforms to hcp suggesting the
gradient of the boundary is positive in the $p-T$ plane.

Whilst the LDL-HDL transition becomes wholly metastable for $r_1\lesssim
1.62$, the lines of density anomalies is nevertheless observable in the
stable liquid region for this value of $r_1$ and indeed a considerable
range of smaller ones. However, since no density anomaly occurs for the
Lennard-Jones potential, it is pertinent to ask how the anomaly
disappears as we approach this limit. To answer this question we have
studied the case $r_1=1.3$ (cf. fig.~\ref{fig:family}), which is much
closer to the LJ limit than the potentials discussed so far. Here we
find that freezing occurs at much higher temperatures than found for our
studies of the range $r_1=1.72-1.62$, no anomalies are seen and there is
no indication of a metastable liquid-liquid transition. It thus appears
that a rapid increase in the freezing temperature occurs with decreasing
$r_1$ (as already hinted at in fig.~\ref{fig:pd}). As a result the
stable solid region engulfs the temperature range in which the anomalies
would otherwise occur. This occurs despite the fact that the maximum
temperature attained by the line of anomalies appears to increase slowly
as $r_1$ decreases (cf. fig.~\ref{fig:anomalies}). We estimate that the
anomalies are lost for $r_1\lesssim 1.4$.

\section{Discussion and conclusions}

To summarize, previous simulation work on ramp potentials
\cite{JAGLA01,WILDING02} has been confined to the situation in which the
LLCP occupies the stable fluid region. Here the LL phase boundary has a
positive gradient in the pressure-temperature plane of the phase
diagram.  We have shown that by judicious choice of ramp parameters, one
can render the LLCP metastable with respect to freezing to a crystalline
solid of density lower than that of the liquid. A line of density maxima
emanates from near the metastable LLCP and extends well into the {\em
stable} fluid region. The line of density maxima bends back in the $p-T$
plane as pressure is reduced. Furthermore (and in contrast to its stable
counterpart), the gradient of the metastable LL phase boundary is {\em
negative}. All these features are in qualitative agreement with the
results of simulations of water, and as such, our results lend
substantial weight to the `second critical point' hypothesis for water.

It is probably fair to say that there is currently no clear picture
regarding the factors controlling (i) the existence or otherwise of
density anomalies in terms of the form of the interparticle potential;
and (ii) the detailed relationship between any such line of anomalies
and the LL phse boundary.  In situations where a line of density maxima
exists, this is thought to be a sufficient, but not a necessary
condition for a LLCP to occur, at least for supercooled states
\cite{SCIORTINO03}. However, it remains unclear why shoulder potentials
exhibit a LL transition, but no density anomaly, while ramp potentials
exhibit both. As regards the locus of the line,  thermodynamic
considerations limit the number of ways in which it can terminate
\cite{SPEEDY82,DEBENEDETTI86}; specifically it must either intersect a
spinodal or transform smoothly into a line of density minima. For the
family of ramp potentials studied in the present work, the line of
density anomalies was always found to approach the LLCP at their high
pressure end. Indeed the same appears to be true for a number of other
distinct models exhibiting LL transitions \cite{AT}, although there are
yet other models where the intersection appears to occur at a point
further down the LL boundary \cite{OTHERS,SCIORTINO03}. We have recently
obtained preliminary results for the ramp potential which may potentiall
shed some light on this matter. Specifically we find that if the
interation range is increased to values greater that those studied here,
the line of density anomalies detaches from the LLCP; its intersection
with the LL boundary then occurs at sub-critical temperatures and in a
region of negative pressure.  We hope to report on this finding in
greater detail in a future publication.


\begin{thebibliography}{50}
\bibitem{NOTE0} We note that one should expect a kink in the freezing
boundary at the point where it intersects the LDL-HDL coexistence line.
However the methods used in this work were unable to resolve this
feature.
\end{thebibliography}
\end{document}